\documentclass[preprint,12pt]{elsarticle}

\usepackage{amssymb}
\usepackage{amsmath}

\newcommand{\be}{\begin{equation}}
\newcommand{\ee}{\end{equation}}
\newcommand{\ba}{\begin{eqnarray}}
\newcommand{\ea}{\end{eqnarray}}

\begin{document}

\begin{frontmatter}



\title{Transport coefficients of hot and dense matter in
relativistic heavy-ion collisions} 


\author{Sukanya Mitra}
\ead{sukanya.mitra@niser.ac.in}
\address{School of Physical Sciences, National Institute of Science Education and Research, An OCC of Homi Bhabha National Institute, Jatni-752050, India.}


\begin{abstract}
The transport coefficients are known as the measure of system interactions, as well as the dynamical input of the hydrodynamic evolution equations of an expanding system created in the relativistic heavy ion collisions. In the current analysis, they have been evaluated for strongly interacting quark-gluon plasma system as well as hot hadronic media. The medium effects for an interacting QCD system have been introduced through a quasiparticle model. The same for a hadronic system has been captured via the in-medium self-energy correction using thermal field theory. The resulting behavior of the transport coefficients shows significant modification with respect to the estimations made in vacuum or using ideal equation of state.
\end{abstract}


\end{frontmatter}

\section{Introduction}
In the last few decades, the sub-nucleonic world of partonic substructures (quarks and gluons) has been studied with greater precision by exploring a deconfined state of the nuclear matter at experimental facilities like RHIC (BNL, USA) and LHC (CERN, Geneva), known as the strongly coupled Quark-Gluon-Plasma (QGP) \cite{QGP-exp}. Relativistic hydrodynamics has played an important role to understand the data from these experiments \cite{Heinz:2013th} and providing a viable description of the collective dynamics of the produced matter. Opposing the preconceived notion that it is either a weakly interacting gas or like a viscous honey, the QGP behaves like a nearly ideal fluid, albeit including non-negligible dissipative corrections. Recently, the observation of a large elliptic flow ($v_2$) of hadrons in $200$ GeV Au-Au collisions at the RHIC could be explained quantitatively using a small but finite value of the shear viscosity over entropy density $(\eta/s\sim 2\times (1/4\pi))$ \cite{Luzum:2008cw}. Moreover, these dissipative quantities, commonly quantified by the ``transport coefficients'' operate as the input parameters in the hydro evolution equations and control the space-time behavior of the thermodynamic quantities critically.
Furthermore, they turn out to be useful signatures of the dynamical interaction and also indicate the phase transition in the created medium at RHIC and LHC \cite{Niemi-Denicol}. The value of $\eta/s$ of quantum chromodynamic (QCD) matter cannot be constant \cite{Prakash:1993bt}, it is expected to display a strong temperature dependence with a minimum around the phase transition or crossover region of QGP to hadronic matter transition \cite{Csernai:2006zz}. These findings serve as a prime motive to study the transport coefficients in the hot and dense matter created at relativistic heavy-ion collisions.

Though the hydrodynamic equations may be derived from
entropy considerations using the second law of thermodynamics, a coarse-grained microscopic approach is necessary in order to extract the transport coefficients, namely the coefficients of shear and bulk viscosity, thermal and charge conductivity, and the relaxation times of the corresponding fluxes. In this context, the covariant kinetic theory involving the relativistic Boltzmann transport equation has extensively served as the underlying microscopic theory to estimate the transport coefficients of the relativistic imperfect fluids. The existing studies of transport properties in this approach cover a significant amount of literature extending both in strongly interacting QGP \cite{QGPViscosity} and hot hadronic matter \cite{HadronViscosity}.

In the current analysis, the transport coefficients have been estimated from the relativistic transport equation for both strongly interacting QGP and hot hadronic medium. For the hot hadronic system, the tools from thermal field theory have been employed to include medium effects. For strongly interacting QGP the method of quasiparticle model has been implemented to incorporate the effects of hot QCD equations of state (EOS).
The manuscript is organized as follows. Section \ref{framework} provides the formal framework for estimating transport coefficients in the gradient expansion technique and moment method. In section \ref{hadron} and \ref{QGP} the transport coefficients in hot hadronic matter and interacting QGP matter have been evaluated respectively. Finally, in section \ref{conclusion} the manuscript is closed with a summary and a possible outlook.

\section{Formal framework}
\label{framework}
In relativistic kinetic theory, the estimation of transport coefficients is basically done by comparing the macroscopic and microscopic definitions of thermodynamic fluxes (sometimes known as thermodynamic flows as well that tends to smooth out the non-uniformities created by thermodynamic forces).
In order to do so, first we identify the thermodynamic fields such as the particle four-flow and energy momentum tensor decomposed in an equilibrium and non-equilibrium part respectively given as follows,
\begin{align}
N^{\mu}(x)&=N^{\mu}_0+\delta N^{\mu}=n_0u^{\mu}+V^{\mu}=\int\frac{d^3p}{(2\pi)^3p^0}p^{\mu}(f_0+\delta f)~,\nonumber\\
T^{\mu\nu}(x)&=T^{\mu\nu}_0+\delta T^{\mu\nu}=\varepsilon_0 u^{\mu}u^{\nu}-(P_0+\Pi)\Delta^{\mu\nu}+\left[W^{\mu}u^{\nu}+W^{\nu}u^{\mu}\right]+\pi^{\langle\mu\nu\rangle}~,\nonumber\\
&=\int\frac{d^3p}{(2\pi)^3p^0}p^{\mu}p^{\nu} (f_0+\delta f)~.
\label{flow}
\end{align}
We define, $n=$ particle number density, $\varepsilon=$ energy density, $T=$ temperature, $P=$ pressure, $u^{\mu}=$ hydro four-velocity, $V^{\mu}=$ diffusion flow, $W^{\mu}=$ heat flow, $\Pi=$ bulk viscous flow, $\pi^{\mu\nu}=$ shear viscous flow, $p^{\mu}=$ particle four-flow, $x^{\mu}=$ space-time coordinate, $f=$ single particle distribution function. We identify the subscript $0$ refers to equilibrium and $\delta$ refers to non-equilibrium part of the associated quantities.
It is the out of equilibrium particle distribution $\delta f$ in Eq.\eqref{flow} that needs to be solved from the relativistic Boltzmann transport equation which is given as the following \cite{Degroot-Cercignani},
\begin{align}
 p^{\mu}\partial_{\mu} f(x,p) =C[f]~,
 \label{RBT}
\end{align}
with the collision term explicitly constraining the reaction rate (dynamical input) via differential scattering cross section, $W=\frac{s}{2}\frac{d\sigma}{d\Omega}(2\pi)^6 \delta^4(p+k-p'-k')$. In order to solve \eqref{RBT} the following two methods are generally opted.

\subsection{Gradient expansion technique}
In this technique we adopt the well known Chapman-Enskog (CE) approximation, which is an iterative method where the out of equilibrium distribution function is successively estimated through order by order gradient correction. In a nutshell it can be explained as follows. The associated length scale of the left hand side of transport equation \eqref{RBT} is of the dimension of spatial non-uniformities $\sim L$, where the associated length scale of the right hand side collision term is the mean free path $\sim\lambda$. The ratio is called the Knudsen number = Kn=$\lambda/L$ and Kn$<1$ denotes the hydrodynamic regime. In CE method the out of equilibrium distribution function is expanded order by order in term of the expansion parameter Kn such as,
$
f = f_0 + \text{(Kn)} f_1+(\text{Kn})^2 f_2+\cdots~,
$ $f_n$ being $n^{th}$ order correction.
In this manner the first-order CE approximation turns out to be,
\begin{align}
p^{\mu}\partial_{\mu}f_0(x,p)=-{\cal{R}[\phi]}~,~~~~~~~~
f_1=f_0+f_0\left(1\pm f_0\right)\phi~.
\end{align}
 with the linearized collision term over first order distribution correction $\phi$,
 \begin{align}
 {\cal{R}}[\phi]=&\int \frac{d^3k}{(2\pi)^3k^0} \frac{d^3p'}{(2\pi)^3p^{'0}} \frac{d^3k'}{(2\pi)^3k^{'0}}f(x,p)f(x,k)\{1\pm f(x,p')\}\{1\pm f(x,k')\}\nonumber\\
 &\times \left[\phi(x,p)+\phi(x,k)-\phi(x,p')-\phi(x,k')\right] W\left(p,k|p',k'\right)~.
\end{align}
Solving this for first order distribution correction and putting in \eqref{flow} we obtain the expression for the transport coefficients like shear and bulk viscosity, and thermal conductivity respectively as follows:
\begin{align}
 &\frac{\zeta}{T^3}=\frac{\left\{\frac{z^2}{3}I_2 +\frac{1}{T}\left(\frac{\partial P}{\partial n}\right)_{\epsilon} I_3 +
 \left(\left(\frac{\partial P}{\partial\epsilon}\right)_n-\frac{1}{3}I_4\right)\right\}^2}
{\left[\tilde{E}_p^2,~\tilde{E}_p^2\right]}~,\nonumber\\
 &\frac{\lambda}{T^2}=-\frac{\left\{J_2-\left(\frac{\epsilon + P}{nT}\right)J_1\right\}^2}
 {{\left[\tilde{E}_p\tilde{p}^{\langle\mu\rangle},~\tilde{E}_p\tilde{p}_{\langle\mu\rangle}\right]}}~,~~~~
 \frac{2\eta}{T^3}=\frac{\left\{ K_0  \right\}^2}
 {\left[\tilde{p}^{\langle\mu}\tilde{p}^{\nu\rangle},~\tilde{p}_{\langle\mu}\tilde{p}_{\nu\rangle}\right]}~.
\end{align}
with, $[F,G]\sim W$, such that transport coefficients are inversely proportional to the medium interaction. $I_n,J_n,K_n$ are the scalar, vector and tensor moments of particle momenta.
For full derivation and detailed expressions, see \cite{Mitra:2012jq,Mitra:2013gya,Mitra:2014dia,Gangopadhyaya:2016jrj,Mitra:2017sjo,Mitra:2018akk}.

\subsection{Moment method}
The Grad's 14 moment method solves the out of equilibrium distribution function in its entirety rather than order by order. The transport equation \eqref{flow} in this approach turns out to be,
\begin{align}
p^{\mu}\partial_{\mu}f_0+f_0(1\pm f_0)p^{\mu}\partial_{\mu}\phi+\phi p^{\mu}\partial_{\mu}f_0=-{{\cal L}}[\phi]~.
\label{Grad}
\end{align}
Here, $\phi$ is expressed as a sum of scalar products of tensors formed from $\tilde{p}^{\mu}=p^{\mu}/T$ and tensors functions of $x_{\mu}$ as follows,
\begin{align}
\phi=A(x,\tilde{E_p})+B^{\mu}(x,\tilde{E_p})\tilde{p}_{\langle\mu\rangle}+C^{\mu\nu}(x,\tilde{E_p})\tilde{p}_{\langle\mu}\tilde{p}_{\nu\rangle}~,
\end{align}
with the coefficients $A,B^{\mu},C^{\mu\nu}$ further expanded in a power series of $\tilde{E_p}=(p\cdot u)/T$ up to first non-vanishing contribution to the irreversible flows as,
\begin{align}
A=\sum_{s=0}^{2}A^{s}(x)\tilde{E}_p^{s}~,~~~
B^{\mu}=\sum_{s=0}^{1}\{B^{s}(x)\}^{\mu}\tilde{E}_p^s~,~~~
C^{\mu\nu}=\{C^{0}(x)\}^{\mu\nu}~.
\end{align}
Eventually the unknown coefficients are estimated in terms of the dissipative flows as, $A^1, A^2 ,A^3 \sim \Pi,~~\{B^{0}\}^{\mu},\{B^{1}\}^{\mu}\sim W^{\mu},~~\{C^{0}\}^{\mu\nu} \sim \pi^{\mu\nu}$, and $\phi$ is expressed as a linear combination of the thermodynamic flows.
Now all we need is to take the moments of Eq.\eqref{Grad},
where zeroth moment gives conservation of particle four-flow - $\partial_{\mu}N^{\mu}=0$, first moment gives conservation of energy-momentum - $\partial_{\mu} T^{\mu\nu}=0$, and higher moment gives hydrodynamic evolution equations, $\rho^{\mu_1\mu_2\cdots} + \tau_{\rho} D\rho^{\mu_1\mu_2\cdots}= \text{first order terms} $, where $\tau_\rho$ is the relaxation time of dissipative flow $\rho^{\mu_1\mu_2\cdots}$. For detailed derivation of hydro equations along with the expressions of transport coefficients see \cite{Mitra:2015yaa,Mitra:2019jld}.

\section{Transport coefficients in hot hadronic matter}
\label{hadron}
\begin{figure}[h]
\centering
\includegraphics[scale=0.3]{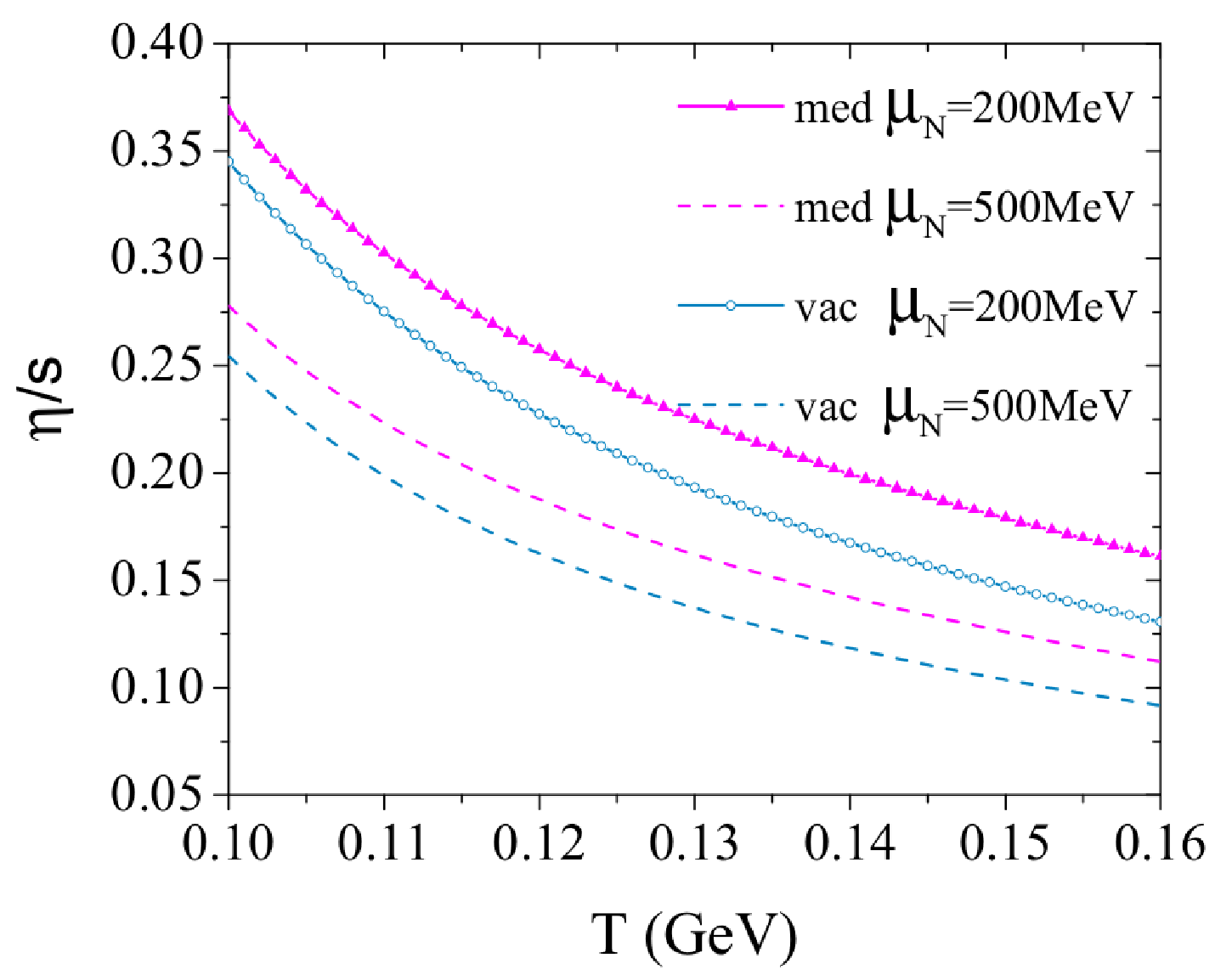}
\caption{$\eta/s$ as function of $T$ in hot hadronic medium.}
\label{etabys}
\end{figure}
We introduce the medium effects in hadronic cross section (actually in the propagator $D_{\mu\nu}$) via one loop self energy correction ($\Pi^{\mu\nu}$) using the thermal field theory as, $D_{\mu\nu}=D_{\mu\nu}^{(0)}+D_{\mu\rho}^{(0)}\Pi^{\rho\lambda}D_{\lambda\nu}$. Because of the medium effects at non-zero temperature, the additional scattering and decay processes increase the decay width and reduce the cross section with respect to vacuum \cite{Mitra:2012jq,Mitra:2013gya,Mitra:2014dia,Gangopadhyaya:2016jrj}. As a result, the temperature dependence of transport coefficients enhances in the presence of the thermal medium. In Fig.\ref{etabys} the shear viscosity over entropy density ratio $\eta/s$ has been plotted as a function of $T$ which shows medium effects clearly enhance the value of $\eta/s$ at both values of nucleonic chemical potential. Hence, the effects of a thermal medium on the temperature dependence of transport coefficients are clearly depicted.

\section{Transport coefficients in QCD matter}
\label{QGP}
\vspace{0.3cm}
\begin{figure}[h]
\centering
\includegraphics[scale=0.3]{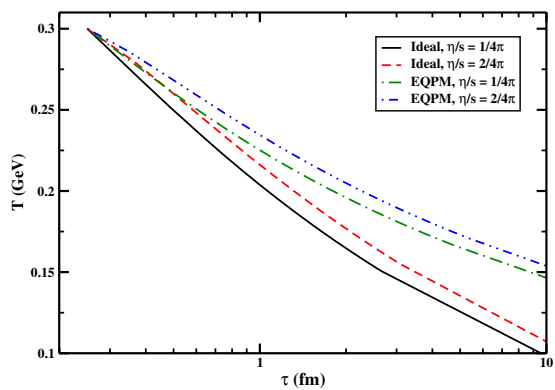}
\caption{Evolution of temperature for different $\eta/s$ ratios with and without the EQPM.}
\label{Ttau}
\end{figure}
\vspace{0.2cm}
\begin{figure}[h]
\centering
\includegraphics[scale=0.3]{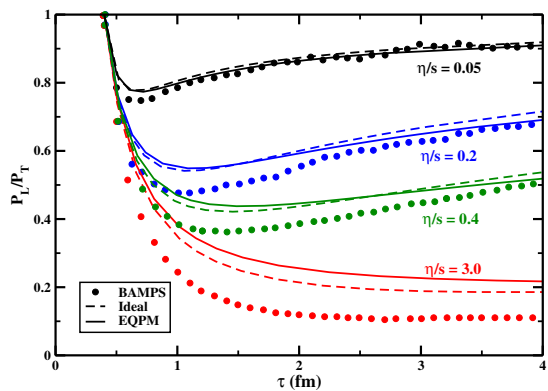}
\caption{Evolution of $P_L/P_T$ with and without the EQPM.}
\label{PLPT}
\end{figure}
In order to consider the strong interaction effects of a QCD medium, we take recourse of a quasiparticle model called the effective fugacity quasiparticle model (EQPM) \cite{Mitra:2017sjo,Mitra:2018akk}. The idea is that
the properties of elementary particles become “dressed” by medium interactions and they are called the quasiparticles. Here the single particle distribution function is described as,
\begin{align}
 f_{k}= {z_{k}(T)\exp\big[\frac {E_{p_k}}{T}-\frac{\mu_{B_k}}{T}\big]}/{\bigg(1\mp z_{k}\exp\big[\frac {E_{p_k}}{T}-\frac{\mu_{B_k}}{T}\big]\bigg)}~,
\end{align}
with energy dispersion $\omega_{p_k}=E_{p_k}+\Delta_k,~~~\Delta_k=T^2\partial_T ln(z_k)$ where the factor $z_k$ is fixed from the equations of state of the (2+1)-flavor lattice QCD simulations at physical quark masses \cite{HotQCD:2014kol}. Using this model, the second order hydro equations for a 1+1 boost invariant system is given by \cite{Mitra:2019jld},
\begin{align}
 \frac{d\epsilon}{d\tau}=-\frac{\epsilon+P}{\tau}+\frac{\phi}{\tau}~,~~~~~~
\frac{d\phi}{d\tau}=-\frac{\phi}{\tau_{\pi}}+\frac{2}{3}\frac{1}{\tau}\frac{2\eta}{\tau_{\pi}}-{\lambda}\frac{\phi}{\tau}~,
\end{align}
with $\phi=-\tau^2 \pi^{\eta_s \eta_s}$ and $\lambda=\frac{1}{3}\tau_{\pi\pi}+\delta_{\pi\pi}$.
In Fig.\ref{Ttau}, the proper time evolution of the temperature has been shown for different $\eta/s$ values with and without the EQPM. The effect of the EQPM for each $\eta/s$ ratio is distinct from the ideal ones, which is more prominent at larger times, i.e., for smaller temperatures.
In Fig.\ref{PLPT}, the pressure anisotropy $P_L/P_T=(P-\phi)/(P+\frac{\phi}{2})$ has been plotted as a function of proper time $\tau$ for different $\eta/s$ values. The disagreement between the BAMPS (parton cascade model) and quasiparticle model at large $\eta/s$ is because of the very different equations of state between BAMPS (uses an ultrarelativistic EOS, $P=\varepsilon/3$), and the current quasiparticle model, which uses a lattice EOS.
Thus the effect of a interacting QCD medium is being embedded via the lattice EOS using the EQPM model.

\section{Conclusion and outlook}
\label{conclusion}
Transport coefficients for a strongly interacting system have been evaluated for both deconfined quark-gluon and hadronic regime of relativistic heavy-ion collision.
For quark sector a lattice inspired quasiparticle model has been implemented to account for the medium interactions. For hot hadronic system the self energy correction of the in-medium propagator has been utilized for the same. The results depict that the finite temperature medium effects, which include the interaction processes, have significantly affected the temperature dependence of transport coefficients in both the regime which can be useful for future hydrodynamic simulations.

\section{Acknowledgements}
For the ﬁnancial support I acknowledge the Department of Atomic Energy, India.

\end{document}